\documentclass{ws-procs975x65}

\newcommand{\Mpl}{M_{\rm pl}}

\renewcommand{\d}{{\rm d}}
\newcommand{\ep}{\epsilon}
\newcommand{\epp}{\epsilon_\phi}
\newcommand{\epc}{\epsilon_\chi}

\newcommand{\fnl}{f_{\mathrm{NL}}}
\newcommand{\tnl}{\tau_{\mathrm{NL}}}
\newcommand{\gnl}{g_{\mathrm{NL}}}
\newcommand{\be}{\begin{equation}}
\newcommand{\ee}{\end{equation}}
\newcommand{\bea}{\begin{eqnarray}}
\newcommand{\eea}{\end{eqnarray}}
\newcommand{\ds}{\displaystyle}

\begin{document}

\title{\uppercase{
Conditions for observable bi and tri-spectra in two-field slow-roll inflation}
}

\author{JOSEPH ELLISTON}

\address{School of Physics and Astronomy, Queen Mary University of London,\\
London E1 4NS, UK\\
E-mail: j.elliston@qmul.ac.uk
}

\begin{abstract}
We find constraints on inflationary dynamics that yield a large local bispectrum and/or trispectrum during two-field slow-roll inflation. This leads to simple relations between the non-Gaussianity parameters, simplifying the Suyama--Yamaguchi inequality and also producing a new result between the trispectrum parameters $\tnl$ and $\gnl$. 
\end{abstract}

\keywords{non-Gaussianity; bispectrum; trispectrum.}

\bodymatter

\section{Background and motivation} \label{sec:introduction}

Observational constraints on the statistics of the primordial curvature perturbation provide a powerful test of inflation. Multi-field inflation is a well-motivated scenario that may be observationally distinguished from single-field inflation by the generation of local non-Gaussianity during the superhorizon evolution of perturbations\cite{Maldacena:2002vr,Lyth:2005fi}.

In the absence of a unique inflationary model, a key task at present is to understand the predictions of \emph{classes} of inflationary models, which observational data may either rule out or constrain. In this short note we derive the types of inflationary dynamics that generate a large non-Gaussianity in two-field slow-roll inflation, which represents the simplest multi-field scenario. Further details may be found in Ref.~\refcite{Elliston:2012wm}. Our work is an extension of earlier bispectrum work by Byrnes {\it et al.}\cite{Byrnes:2008wi} which we review, simplify and then extend to the trispectrum. 

The local non-Gaussianity parameters $\fnl$, $\tnl$ and $\gnl$ are defined from the three and four-point correlators of the primordial curvature perturbation on uniform density hypersurfaces as
\begin{eqnarray}
\label{spectra}
\langle\zeta_{\mathbf k_1}\,\zeta_{\mathbf k_2}\,
\zeta_{\mathbf k_3}\rangle &\equiv& (2\pi)^3 \delta^3 ( {\mathbf k_1}+{\mathbf k_2}+
{\mathbf k_3}) \frac{6}{5} \fnl \big[ P_\zeta(k_1) P_\zeta(k_2) 
+ \mathrm{2~perms} \big] \,,\nonumber\\
\langle\zeta_{\mathbf k_1}\,\zeta_{\mathbf k_2}\,
\zeta_{\mathbf k_3} \, \zeta_{\mathbf k_4} \rangle &\equiv& (2\pi)^3 \delta^3 ( {\mathbf k_1}+{\mathbf k_2})+
{\mathbf k_3}+{\mathbf k_4}) \bigg\{
\tnl \big[ P_\zeta(k_{13}) P_\zeta(k_3) P(k_4) + \mathrm{11~perms} \big] \nonumber\\
&& \quad+ \frac{54}{25}\gnl \big[ P_\zeta(k_2)  P_\zeta(k_2) P_\zeta(k_4) +\mathrm{3~perms} \big] \bigg\}\,,
\end{eqnarray}
where 
$\langle\zeta_{\mathbf k_1}\,\zeta_{\mathbf k_2}\rangle \equiv (2 \pi)^3 \delta^3 ( {\mathbf k_1}+{\mathbf k_2})) P_\zeta$ defines the power spectrum and 
${\mathbf k}_{ij} = {\mathbf k}_{i} + {\mathbf k}_{j}$. We consider inflation driven by two canonical and minimally coupled scalar fields $\phi$ and $\chi$, self-interacting through a potential $W(\phi,\chi)$. Slow-roll requires that the following potential slow-roll parameters are all much smaller than unity:
\be
\epsilon_i = \ds{\frac{\Mpl^2}{2} \frac{W_{,i}^2}{W^2} } \,,\qquad 
\epsilon = \ds{\sum_{i=1}^2 \epsilon_i}\,, \qquad
\eta_{ij} = \ds{{\Mpl^2} \frac{W_{,ij}}{W}} \,, \qquad 
\xi^2_{ijk} = \ds{{\Mpl^3} \sqrt{2 \ep} \frac{W_{,ijk}}{W}}\,,
\label{eq:sr_parameters}
\ee
where $\{i,j,k\} \in \{\phi,\chi\}$ and a comma denotes partial derivatives. 

Our calculations employ the $\delta N$ formalism, which allows the non-Gaussianity parameters to be determined analytically\cite{Lyth:2005fi,Vernizzi:2006ve}, provided the potential is of either the sum-separable form $W = U(\phi) + V(\chi)$\cite{Vernizzi:2006ve}, 
or the product-separable form $W = U(\phi)V(\chi)$\cite{Choi:2007su}. It is useful to define $\theta$ as the angle of evolution in the $\{\phi$, $\chi\}$ phase space such that 
$\epp =\ep \cos^2 \theta$ and $ \epc = \ep \sin^2 \theta$. Assuming both fields to monotonically decrease, $\theta$ is constrained as $0\leq \theta \leq \pi/2$.

\section{Analytic formulae for non-Gaussianity} 
\label{sec:analytics}

We use the rotated field basis $\{\sigma,s\}$ where $\d s / \d t=0$, such that $\sigma$ and $s$ respectively define the adiabatic and isocurvature directions. Labels `$*$' denote quantities evaluated on a flat hypersurface near horizon exit; quantities without a label are evaluated on a later-time uniform density hypersurface. Following Ref.~\refcite{Elliston:2012wm}, we introduce $\alpha$ such that for product-separable potentials $\alpha = \theta$, whereas for sum-separable potentials $\alpha_* = \theta_*$ and subsequently $\alpha$ follows from $\d \alpha / \d \theta = W \sin^2 2 \theta / (W^* \sin^2 2 \alpha)$. The parameter $\alpha$ also follows from the linear $\delta N$ expressions of Vernizzi and Wands\cite{Vernizzi:2006ve}.\newline
\indent {\bf The $\fnl$ parameter}, after much manipulation, assumes the simple form\cite{Elliston:2012wm}
\be
\begin{array}{l l}
\ds{\frac{6}{5} \fnl} \simeq \ds{f\, \Big[- \eta_{ss}^* +
2 \Omega \, (\eta_{ss} - \ep) \Big]} & ~~~~~~\mbox{-- Sum separable} \vspace{2mm}
\,,\\
\ds{\frac{6}{5} \fnl} \simeq \ds{f\, \Big[- \eta_{ss}^* + 2 \eta_{ss} \Big]} & ~~~~~~\mbox{-- Product separable} \,, \vspace{2mm} \\
f = \ds{\frac{\sin^2 2\alpha}{4 \Lambda^2} (\cos^2 \alpha - \cos^2 \theta^*)^2 \,,}
& ~~~~~~ \Omega = \ds{\frac{W^2}{W_*^2} \frac{\sin^2 2 \theta}{\sin^2 2 \alpha}} \,,  
\label{eq:fnl_direction}
\end{array}
\ee
where $\Lambda = \cos^4 \alpha \sin^2 \theta^* + \sin^4 \alpha \cos^2 \theta^*$. Note that $f>0$ and $0 \leq \Omega \leq 1$. We consistently use `$\simeq$' such that equality holds to excellent precision if the non-Gaussianity is large enough to be observable. Eq.~\eqref{eq:fnl_direction} implies that a necessary (but not sufficient) condition for $|\fnl| > 1$ is a fine-tuning $\theta_* \ll 1$ such that $f \gg 1$.\newline
\indent {\bf The $\tnl$ parameter} simplifies as
\be
\tnl \simeq {\cal C} \left( \frac{6}{5} \fnl \right)^2 \,,
\qquad {\cal C} = \frac{\Lambda}{(\cos^2 \alpha - \cos ^2 \theta^*)^2} \geq 1 \,.
\label{eq:tnl}
\ee
which is valid for both sum and product-separable potentials. This approximate equality is a special case of the Suyama--Yamaguchi inequality\cite{Suyama:2007bg}. \newline
\indent {\bf The $\gnl$ parameter} assumes the forms
\begin{align}
\frac{27}{25} \gnl &\simeq 
\tnl \left( \frac{\eta_{ss}^* -\Omega \, (\eta_{ss}-\ep)}{\eta_{ss}^* -2 \, \Omega \, (\eta_{ss}-\ep)} \right) 
-\frac{6}{5} \fnl (2 \eta_{ss}^* + \Omega \, (\eta_{ss}-\ep))
-g_4 {\xi_{sss}^*}^2 \nonumber \\
& \qquad + g_1\, \Omega^{3/2} \, \Big[\xi_{sss}^2 - 2 \eta_{\sigma s}
(\eta_{ss}+\ep)\Big] -\frac{1}{2} f_1 f \ep^* \eta_{ss}^* \nonumber \\
& \qquad + 4 g_3 \, \Omega \, (\eta_{ss}-\ep) \left( \frac{W}{W^*}\cos 2 \theta
\eta_{ss} - \Omega \cos 2 \alpha (\eta_{ss} - \ep) \right)
\,, \label{eq:ss_gnl} 
\\
\frac{27}{25} \gnl &\simeq
\tnl \left( \frac{\eta_{ss}^* -\eta_{ss}}{\eta_{ss}^* -2\eta_{ss}} \right) 
-\frac{6}{5} \fnl (2 \eta_{ss}^* + \eta_{ss})-g_4 {\xi_{sss}^*}^2 
+ g_1 \Big[\xi_{sss}^2 - 2 \eta_{\sigma s} \eta_{ss}\Big], \label{eq:ps_gnl}
\end{align}
for sum and product-separable potentials respectively, where we have used
\be
\begin{array}{llllrll}
\tau_2 &=& \displaystyle{\frac{\sin 2 \theta^*}{\Lambda^3}(\cos ^8 \alpha \sin ^4 \theta^*-\sin ^8 \alpha \cos ^4 \theta^*)}\,,  &&~~~
g_1 &=& \displaystyle{g_3\sin 2 \alpha } \vspace{2mm}\,,  \\

\tau_3 &=& \displaystyle{\frac{f_1}{2 \Lambda^2}(\cos ^8 \alpha \sin ^2 \theta^*+\sin ^8 \alpha \cos ^2 \theta^*)}\,,  &&
g_2 &=& \displaystyle{g_3\cos 2 \alpha } \vspace{2mm}\,,  \\

g_4 &=& \ds{\frac{1}{4} \left(\tau_3 \sin 2 \theta^* \cos 2 \theta^* -\tau_2 \right)} \,, &&
g_3 &=& \displaystyle{-\frac{f}{2 \Lambda}(\cos^2 \alpha - \cos^2 \theta^*)}\,,
\label{eq:ss_tnl_para}
\end{array}
\ee
and $f_1 = \sin^2 2 \theta_* / (2 \Lambda)$. \newline
\indent{\bf Adiabaticity:} If the dynamics reach an adiabatic limit\cite{Elliston:2011et, Elliston:2011dr} during slow-roll inflation then our sum-separable analytic expressions simplify following the Horizon Crossing Approximation ({\sc hca})\cite{Kim:2006te} as $\Omega \to 0$. Working in this limit and assuming the $g_4 {\xi_{sss}^*}^2$ term may be neglected, we find a new relation between $\tnl$ and $\gnl$ as
\be
\frac{27}{25} \gnl \simeq \tnl \,.
\label{eq:gnl_tnl}
\ee
We have found it {\it very} fine-tuned to generate deviations from this result in the {\sc hca}. 

\section{Interpretation and conclusions}

By plotting the functions such as $f$ appearing in eqs. \eqref{eq:fnl_direction}--\eqref{eq:ps_gnl}, we have verified that the inflationary dynamics that give rise to a large bispectrum parameter $\fnl$ are also capable of producing large values of the trispectrum parameters $\tnl$ or $\gnl$. In all cases, a necessary requirement for a large local non-Gaussianity is that the horizon crossing field velocities are dominated by one of the two fields. For quadratic potentials we find $\gnl$ to be subdominant, whereas more general potentials such as inflection points have $\gnl \sim \tnl$. Under the {\sc hca}, we have generated a new consistency relation \eqref{eq:gnl_tnl} between the trispectrum parameters $\gnl$ and $\tnl$. 

\section*{Acknowledgments}
JE is supported by a STFC studentship. We would like to thank Laila Alabidi, Ian Huston, David Mulryne and Reza Tavakol for their support. 

\bibliographystyle{ws-procs975x65}
\bibliography{gnl}

\end{document}